# Hybrid Poisson and multi-Bernoulli filters

Jason L. Williams
Intelligence, Surveillance and Reconnaissance Division
Defence Science and Technology Organisation, Australia
email Jason.Williams@dsto.defence.gov.au

*Abstract*—The probability hypothesis density (PHD) and multi-target multi-Bernoulli (MeMBer) filters are two leading algorithms that have emerged from random finite sets (RFS). In this paper we study a method which combines these two approaches. Our work is motivated by a sister paper, which proves that the full Bayes RFS filter naturally incorporates a Poisson component representing targets that have never been detected, and a linear combination of multi-Bernoulli components representing targets under track. Here we demonstrate the benefit (in speed of track initiation) that maintenance of a Poisson component of undetected targets provides. Subsequently, we propose a method of recycling, which projects Bernoulli components with a low probability of existence onto the Poisson component (as opposed to deleting them). We show that this allows us to achieve similar tracking performance using a fraction of the number of Bernoulli components (*i.e.*, tracks).

## I. INTRODUCTION

In recent years, random finite sets (RFS) [1] has emerged as coherent approach for inference in problems commonly encountered in tracking, involving unlabelled measurements of a set of objects with unknown cardinality. The probability hypothesis density (PHD) [2] and multi-target multi-Bernoulli (MeMBer) [1,3] filters have been shown to be effective in a variety of tracking problems. This paper demonstrates the benefits associated with combining these two structures in a hybrid Poisson-MeMBer tracker. The work is motivated by the derivation in the sister paper [4,5], which shows that, under the common assumption that the target birth process is Poisson, the full Bayes RFS filter consists of both a Poisson component, and a linear combination of multi-Bernoulli distributions, where the Poisson component represents the distribution of targets that have *never* been detected. We show the benefit of incorporating both of these components in their natural roles, and subsequently show the benefit of extending the Poisson component to represent tracks (*i.e.*, multi-Bernoulli components) with a low probability of existence.

At first glance, it may seem unusual to be maintaining a distribution of targets that have never been detected. The reasons for doing so may be best understood in the context of agile sensors such as phased array radars. Intuitively, if the radar has not observed a region of space for a long period, it is more likely that there will be new targets in that portion of space awaiting first detection. Conversely, if a region has been observed very recently, we expect that it is comparatively unlikely that undetected targets will be present. The mathematical model behind this intuition is that targets arrive (and depart) at a constant rate regardless of whether they are detected. We thus need to capture this arrival and departure of targets even in portions of space we are not presently observing in order to be able to predict the expected number of new targets in the area when it is next observed.

Practically, the distribution of undetected targets is useful for two purposes. Firstly, it can be used for sensor scheduling, steering an agile sensor in order to minimise the expected number of undetected targets. This has been studied recently in [6], and is not considered in this paper. Secondly, it has been shown [4] that the Bayes RFS filter naturally incorporates the distribution of undetected targets into its calculation of the probability that any detection represents a new target. Thus, maintaining a distribution of undetected targets allows the Bayes RFS filter to achieve the best possible track initiation performance, since it utilises the information to anticipate the relative likelihood of a measurement representing a false alarm, a target currently under track, or a new target.

The multi-Bernoulli representation has significant advantages compared to the PHD as it is able to accumulate high confidence in the existence of targets. The major disadvantage is the requirement to maintain a large number of tracks with a small probability of existence in order to achieve adequate track initiation performance. The hybrid Poisson-multi-Bernoulli representation proposed in this paper motivates a true hybrid PHD-MeMBer algorithm which uses the Poisson component not only for undetected targets, but also for tracks with low probability of existence. This hybrid approach is shown to result in improved track initiation performance over the conventional method while maintaining only a small fraction of the number of tracks. A similar concept was proposed recently in [7]; we discuss the relationship with this work in Section V.

### A. Outline and contributions

Section II describes the necessary results from the sister paper [4] (which shows that the Bayes RFS filter naturally incorporates both Poisson and multi-Bernoulli components), concentrating on the role of the Poisson component and its interaction with the multi-Bernoulli component. The remaining sections provide new contributions including:

- Demonstration of the benefit of the hybrid Poisson-multi-Bernoulli representation, in both uniform, stationary problems (Section III-A), and in problems involving agile sensors (Section III-B)

- Principled development of a hybrid PHD-MeMBer algorithm, providing a method for projecting a subset of multi-Bernoulli tracks onto the Poisson component, and an analysis of the distortion that this causes (in turn, providing insight into the recommended existence threshold to use in the projection) (Section IV-A)
- Demonstration of the practical benefit of the hybrid PHD-MeMBer approach (Section IV-B)

The relationship of the proposed method to prior work is described in Section V.

## II. Background

### A. Assumptions and notation

The assumptions and notation include:

- The state of objects is denoted by $x \in \mathcal{X}$ (*e.g.*, position and velocity in two dimensions)
- Targets arrive according to a non-homogeneous Poisson point process (PPP) with intensity $\lambda^{\text{b}}(x)$, independent of existing targets
- Targets depart according to independent, identically distributed (iid) Markovian processes; the survival probability in state $x$ is $P^{\text{s}}(x)$
- Motion for each target is governed by a Markovian process, independent of all other targets; the single-target transition probability density function (PDF) is $f_{t|t-1}(x|x')$
- Each target may give rise to at most one measurement; the probability of detection in state $x$ at time $t$ is $P^{\text{d}}_t(x)$
- Each measurement $z \in \mathcal{Z}$ (*e.g.*, position in two dimensions, or range and azimuth) is the result of at most one target
- False alarms arrive at time $t$ according to a non-homogeneous PPP with intensity $\lambda^{\text{fa}}_t(z)$, independent of targets and target-related measurements
- Each target-derived measurement is independent of all other targets and measurements conditioned its corresponding target; the single target measurement likelihood is $f_t(z|x)$

We denote by $Z_t = \{z_t^1, \ldots, z_t^{m_t}\}$ the measurement set at time $t$, and by $Z^t = \{Z_1, \ldots, Z_t\}$ the measurement history up to and including time $t$. The multiple sensor case may be addressed by performing update steps for each sensor sequentially in between prediction steps.

### B. Review of results from sister paper

The sister paper [4] provides a form of conjugate prior for the tracking problem under the above assumptions. The form involves the union of an independent Poisson process and a linear combination of multi-Bernoulli distributions. A practical algorithm is obtained by approximating the linear combination of multi-Bernoulli distributions as being multi-Bernoulli. The components of this form are:

- a PPP representing undetected targets, with intensity $\lambda^{\text{u}}_{t|t'}(x)$, and
- a series of Bernoulli tracks, $i \in \mathcal{T}_{t'} = \{1, \ldots, n_{t'}\}$,[1]

$$f^i_{t|t'}(X) = \begin{cases} 1 - q^i_{t|t'}, & X = \emptyset \\ q^i_{t|t'} f^i_{t|t'}(x), & X = \{x\} \\ 0, & |X| \geq 2 \end{cases} \quad (1)$$

where $q^i_{t|t'}$ is the probability of existence for the track, and $f^i_{t|t'}(x)$ is the existence-conditioned PDF

The full target distribution can be reconstituted from these components as [1, p386]

$$f_{t|t'}(X) = \sum_{Y \subseteq X} f^{\text{u}}_{t|t'}(Y) f^p_{t|t'}(X - Y) \quad (2)$$

where $f^{\text{u}}_{t|t'}(X)$ is the Poisson distribution of undetected targets:[2] [1, p366]

$$f^{\text{u}}_{t|t'}(X) \propto \prod_{x \in X} \lambda^{\text{u}}_{t|t'}(x) \quad (3)$$

and $f^p_{t|t'}(X)$ is the multi-Bernoulli distribution of targets under track:

$$f^p_{t|t'}(\{x_1, \ldots, x_n\}) \propto \sum_{\alpha} \prod_{i \in \mathcal{T}_{t'}} f^i_{t|t'}(X_{\alpha(i)}) \quad (4)$$

where the sum is over all functions $\alpha : \mathcal{T}_{t'} \to \{0, \ldots, n\}$ such that $\{1, \ldots, n\} \subseteq \alpha(\mathcal{T}_{t'})$ and if $\alpha(i) > 0$, $i \neq j$ then $\alpha(i) \neq \alpha(j)$, and

$$X_{\alpha(i)} \triangleq \begin{cases} \{x_{\alpha(i)}\}, & \alpha(i) > 0 \\ \emptyset, & \alpha(i) = 0 \end{cases}$$

The PPP intensity of undetected targets is predicted in the same manner as a PHD filter [2,4]

$$\lambda^{\text{u}}_{t|t-1}(x) = \lambda^{\text{b}}(x) + \int f_{t|t-1}(x|x') P^{\text{s}}(x') \lambda^{\text{u}}_{t-1|t-1}(x') \mathrm{d}x' \quad (5)$$

and updated in the same way as a PHD if there are no measurements present

$$\lambda^{\text{u}}_{t|t}(x) = [1 - P^{\text{d}}_t(x)] \lambda^{\text{u}}_{t|t-1}(x) \quad (6)$$

We only describe the aspects of the multi-Bernoulli component of the distribution $f^p_{t|t'}(X)$ that relate to the interaction with the undetected target density; a full derivation can be found in [4]. Under the marginal track filter (MTF), a new multi-Bernoulli component $i$ is introduced for each measurement $z$ in each scan. Using the hypothesis $\tilde{a}$ that the measurement $z$ is associated with a pre-existing track, the hypothesis-conditioned distribution is $f^{i,\tilde{a}}_{t|t}(\emptyset) = 1$, *i.e.*, probability of existence is zero. Under the hypothesis $a$ that the measurement is not associated with any previous track, the hypothesis-conditioned distribution is Bernoulli, with

$$q_{t|t}^{i,a} = \frac{\lambda_{t|t-1}^{u}[f_t(z|\cdot)P_t^d]}{\lambda_t^{fa}(z) + \lambda_{t|t-1}^{u}[f_t(z|\cdot)P_t^d]} \quad (7)$$

$$f_{t|t}^{i,a}(x) = \frac{f_t(z|x)P_t^d(x)\lambda_{t|t-1}^{u}(x)}{\lambda_{t|t-1}^{u}[f_t(z|\cdot)P_t^d]} \quad (8)$$

where

$$\lambda_{t|t-1}^{u}[f_t(z|\cdot)P_t^d] \triangleq \int f_t(z|x)P_t^d(x)\lambda_{t|t-1}^{u}(x)\mathrm{d}x$$

is the Poisson intensity of measurements arriving from previously undetected targets. The MTF operates by calculating the marginal probability of each measurement-track association event (considering all joint association hypotheses), and representing the posterior distribution via Bernoulli components, where each is averaged over the marginal association events for that track. In the case of interest, this will simply result in

$$q_{t|t}^{i} = p^i(a)q_{t|t}^{i,a} \quad f_{t|t}^{i}(x) = f_{t|t}^{i,a}(x) \quad (9)$$

where $p^i(a)$ is the marginal probability that the measurement $z$ is not associated with any pre-existing track ($p^i(a)+p^i(\tilde{a}) = 1$). As described in [4], this marginal probability incorporates the weight factor $w_{t|t}^{i,a} = \lambda_t^{fa}(z) + \lambda_{t|t-1}^{u}[f_t(z|\cdot)P_t^d]$; this factor replaces the more common false alarm intensity $\lambda_t^{fa}(z)$ in the standard JPDA association probability calculation. As described in [4], we utilise an efficient approximation of the joint association probabilities, calculated via loopy belief propagation (LBP) [8,9].

---

[1] We leave off conditioning on the measurement set history $Z^{t'} = (Z_1, \ldots, Z_{t'})$ throughout, as this is implicit in the second subscript $f_{t|t'}$.
[2] *i.e.*, targets that have *never* previously been detected

In summary, the interaction between the undetected target component and the multi-Bernoulli, previously-detected target component is

- The probability of existence of a new Bernoulli component started on a measurement is calculated as the ratio between the new target measurement intensity and the false alarm measurement intensity plus the new target measurement intensity (weighted by the association probability). Thus the initial probability of existence will be low if the new target intensity is much lower than the false alarm intensity, and high in the opposite case.
- The kinematic distribution of a new Bernoulli component started on a measurement is proportional to the undetected target intensity multiplied by the state-dependent detection probability and the measurement likelihood. Thus a method is provided for incorporating prior information in the track initialisation.
- In calculation of the marginal association probability, the false alarm intensity in the standard JPDA expression is replaced by the sum of the false alarm intensity and the new target measurement intensity. Accordingly, a measurement in an area of high new target intensity is less likely to be associated with a previously existing track. This provides a principled approach for initiation of new targets in the vicinity of existing tracks.

## III. MODELLING UNDETECTED TARGETS

In this section, we explore two simple cases in which there is an advantage associated with modelling a distribution of undetected targets. To commence, we consider the simplest case, with uniform birth, death and probability of detection. We subsequently consider an adaptive, non-homogeneous case, which is representative of a UAV sensor.

### A. Example: Improved initialisation in uniform/stationary case

The first problem that we consider is the most common case studied in academia, in which the probability of detection, birth probability and death probability are all uniform over the region of interest and constant with time. Even under these assumptions, we show that it is advantageous to represent a distribution of undetected targets, which can be conveniently parameterised via a single scalar value.

The advantage comes from the initialisation of the tracker. The most common initialisation does not incorporate any tracks, and effectively assumes that no targets are present when the system commences operation, and that targets gradually arrive from that time in accordance with the birth intensity. However, the fact that the tracker has not been operating generally does not imply that targets are not present.

The Poisson distribution of undetected targets provides a method for incorporating prior information on the expected number of targets present, which is taken into account in calculating the existence probability of new Bernoulli components as new measurements are received. Perhaps the most natural initialisation of the distribution of undetected targets is the steady state distribution when the sensor is not active. Under the uniform and constant in time assumptions, this is found by repeatedly applying the prediction equation:[3]

$$\lambda_{t+1|t}^{u} = \lambda^b + P^s \lambda_{t|t-1}^{u}$$

Clearly, this will eventually reach steady state with $\lambda_{ss}^{u} = \frac{\lambda^b}{1-P^s}$. Thus $\lambda_{0|0}^{u} = \lambda_{ss}^{u}$ is the natural initialisation of the tracker (with $\mathcal{T}_0 = \emptyset$, i.e., no multi-Bernoulli components).

We illustrate the advantage that this provides through a simple scenario in which targets arrive uniformly in the region $[-100, 100]^2$ and velocity region $[-1, 1]^2$, such that the hypervolume of the state space is $|\mathcal{X}| = 200^2 \times 2^2$. The state is $x = [p_x, v_x, p_y, v_y]^T$. We set the birth intensity to $\lambda^b = 0.05/|\mathcal{X}|$ (i.e., on average a new target arrives every 20 time steps) and the survival probability to $P^s = 0.999$ (i.e., on average targets survive 1000 time steps; targets and hypotheses are removed when they depart the position region $[-100, 100]^2$). Thus $\lambda_{ss}^{u} = 50/|\mathcal{X}|$ (i.e., the expected number of targets present is 50). We study a problem involving a low probability of detection, $P^d = 0.3$, such that several measurements are required in order to accrue confidence in the probability of existence of a track. Measurements of position are received, *i.e.*, $p(z|x) = \mathcal{N}\{z; Hx, R\}$, where $R = I$ and

$$H = \begin{bmatrix} 1 & 0 & 0 & 0 \\ 0 & 0 & 1 & 0 \end{bmatrix}$$

---
[3]We also require the transition kernel to satisfy (approximately, at least) $\int_{\mathcal{X}} f_{t|t-1}(x|x')\mathrm{d}x' = 1$.

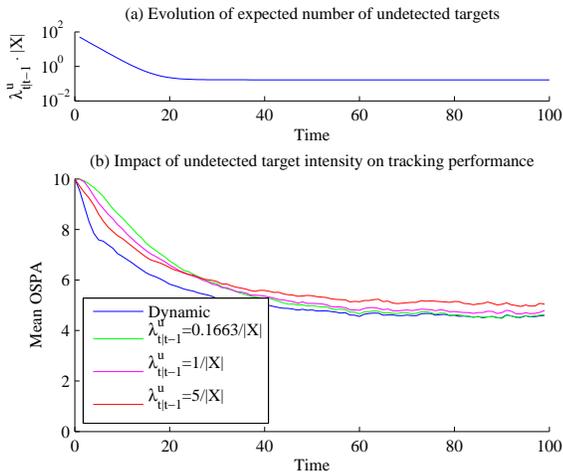

Fig. 1. Performance benefit achieved by maintain undetected target intensity. Upper figure (a) shows evolution of total expected number of undetected targets over time, commencing from the steady state value (50) when the sensor is not operating to the steady state value 0.1663 when the sensor is operating. Lower figure shows the tracking performance (measured by MOSPA) achieved by dynamically estimating the expected number of undetected targets, versus assuming different fixed intensities.

We consider the measurement space to be $\mathcal{Z} = [-100, 100]^2$. The false alarm intensity is $\lambda^{\text{fa}} = 10/|\mathcal{Z}|$ on $\mathcal{Z}$ (*i.e.*, the expected number of false alarms per scan is 10).[4] For the calculation of the updated kinematic distribution of a track upon initialisation, we approximate $\lambda^{\text{u}}_{t|t-1}(x) \approx (\lambda^{\text{u}}_{t|t-1} \cdot |\mathcal{X}|)\mathcal{N}\{x; 0, P\}$ where $P = \text{diag}[100^2/3, 1/12, 100^2/3, 1/12]$. This approximation matches the mean and covariance of the uniform distribution on $\mathcal{X}$. the target kinematics follow the widely-used model with velocity as a random walk, with diffusion $q = 0.01$. The multi-Bernoulli tracker is described in [4]; it outputs all tracks with a probability of existence of at least 0.8, and for which the trace of the covariance is less than 10.

The simulation runs for 100 time steps. The expected number of undetected targets is shown in Fig. 1(a) as a function of time. As expected, the initial number of undetected targets decays from a large value initially, to steady state operation by approximately $t = 30$. In Fig. 1(b), we show the average performance as measured by the mean optimal subpattern assignment (MOSPA) metric [10] (measured against the full state vector), with $p = 2$ and $c = 10$. We compare the proposed dynamic method for estimation of the undetected target intensity to alternatives which hold the intensity of undetected targets constant at $\lambda^{\text{u}}_{t|t-1} \cdot |\mathcal{X}| = 5$, 1, and 0.1663 (the steady state value with the measurement process operating). The figure shows that the method incorporating the dynamic estimate of undetected target intensity outperforms the steady state alternatives throughout the simulation. As expected the filter assuming the steady state intensity performs well at the end of the simulation (when the steady state condition prevails), whereas the filters assuming higher intensity perform better (than other steady state alternatives) during the initial transient. Dynamically modelling the expected number of undetected targets can speed the process of initially acquiring tracks in a scene without trading off steady state performance.

*B. Example: Adaptive initiation for moving sensor*

In the more general (and rarely studied) case in which the distribution of undetected targets is non-homogeneous, an appropriate representation must be chosen. We propose discretising the state space in order to represent this distribution via a grid filter. Grid-based representations (*e.g.*, [11]) have fallen out of vogue for most estimation applications as they are generally inefficient. In particular, in any problem where the probability distributions are peaked, grid representations result in large computing resources being spent on the vast majority of cells with near-zero probability density. In comparison, the preferred sample-based methods allow computing resources to be focussed on the region with significant probability mass. However, in the present application, the Poisson intensity function of undetected targets is, by nature, both diffuse and smooth. Thus a grid-based representation is efficient (since most grid points will have comparable intensity), and, furthermore, a coarse discretisation suffices, such that the grid-based method is tractable. Conversely, sample-based methods using any reasonable number of samples would be likely to exhibit poor sample coverage in the vicinity of many cells.

To demonstrate the more general case, we consider a problem in which the sensor is moving, and the probability of detection is zero outside of a cone $45°$ either side of the sensor's heading. The sensor trajectory is shown in Fig. 2 in blue (it commences at $(-20, -20)$, heading directly down the map). All other parameters remain identical to Section III-A, except that there are two grid cells (shown as magenta stars in Fig. 2) in which the birth rate is equivalent to 20% of the birth rate in the entire region in the previous case (*e.g.*, representing airports). False alarms in regions where the detection probability is zero are naturally discarded by the update equations.

The discretisation used to represent the intensity of undetected targets centres cells on $\{-100, -96, \ldots, 96, 100\}$ in position dimensions, and $\{-1, -0.6, -0.2, 0.2, 0.6, 1\}$ in velocity dimensions, for a total number around 94,000 cells. The transition kernel is obtained by offline Monte Carlo simulation, drawing large number of samples uniformly within the cells, simulating the continuous dynamics process, and observing in which cell they arrive. The survival probability and birth density are modified at the region boundary to achieve a uniform steady state intensity (in the absence of the "airport" entry points in the birth intensity). Fig. 2 shows a snapshot of the scenario, at time $t = 110$ (the scenario duration is 140 time steps). The background colour of the figure shows the undetected target intensity (summing over velocity dimensions); lighter shades denote lower intensity, while darker shades denote higher intensity. The region presently being observed has the lowest intensity while the area that has not been observed for the longest duration has the highest intensity (as expected). Additionally, intensity is increased near the region boundary, modelling entry of new objects into the region.

---
[4]The measurement space is chosen to be larger than the birth region in order to avoid targets departing the portion of the space where false alarms exist. The tracker assumes that both the undetected target intensity and false alarm intensity are constant throughout the space.

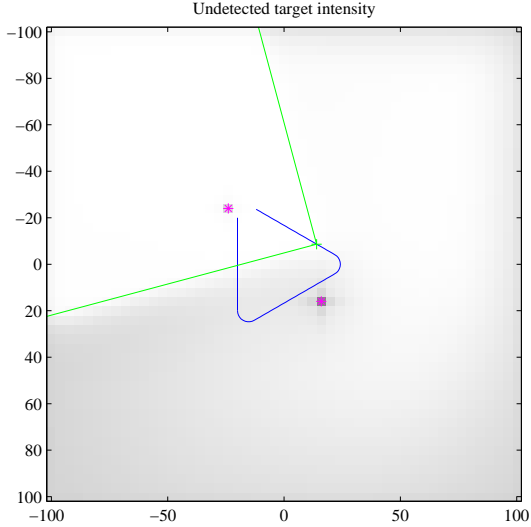

Fig. 2. Undetected target intensity at time $t = 110$ (darker shades denote higher undetected target intensity). Sensor trajectory is show in blue, and field of view at $t = 110$ is shown in green ($P^d = 0.3$ within the field of view, and zero outside).

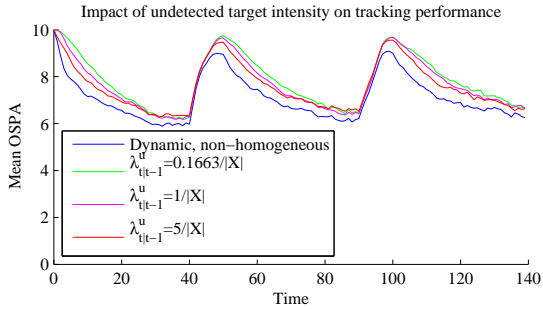

Fig. 3. Performance benefit obtained by maintaining an adaptive, non-homogeneous estimate of undetected target intensity for case of a moving, manoeuvring sensor. Blue line shows performance with dynamic estimate, versus alternatives assuming different fixed, uniform intensities.

The tracking performance for the scenario is shown in Fig. 3. The diagram again shows the MOSPA measure of tracking performance, this time counting only those targets that are within the sensor coverage region. The diagram shows that, each time the sensor manoeuvres, a performance transient occurs which is similar to resetting the tracker. The proposed method which maintains a dynamic, non-homogeneous estimate of undetected target intensity is exhibits an significant improvement in its ability to adapt to these transients. Again, the consequence of this is faster track initiation during this transient, without compromising the rate of false track occurrence at other times.

## IV. RECYCLING: HYBRID PHD-MEMBER

### A. Derivation of method

Recall again (from Section II-B) the form that the MTF maintains from one time interval to the next. In probability generating functional (PGFl) [1, p371] form, the overall target distribution (incorporating both undetected and previous detected targets) is [4]

$$G_{t|t}[h] \propto \exp\{\lambda^{\mathrm{u}}_{t|t}[h]\} \cdot \prod_{i \in \mathcal{T}_t} G^i_{t|t}[h] \qquad (10)$$

where

$$\lambda^{\mathrm{u}}_{t|t}[h] = \int h(x) \lambda^{\mathrm{u}}_{t|t}(x) \mathrm{d}x$$

$$G^i_{t|t}[h] = 1 - q^i_{t|t} + q^i_{t|t} \int h(x) f^i_{t|t}(x) \mathrm{d}x$$

The PGFl form makes clear that this is a union of independent components: a Poisson point process component, and a series of independent multi-target Bernoulli components.[5] Since a new track must be created for every new measurement received, tracks will inevitably need to be deleted from the system. The system designer must trade off system performance against computational complexity in constructing this mechanism. The major source of low probability of existence tracks is initiation. If the probability of detection is low and the false alarm rate is high, then many tracks will be created on false alarms, and it will take a large number of measurement scans to reduce the probability of existence of the tracks to the point where they can be excluded with confidence.

One possibility which arises from the form of (10) is to approximate individual tracks as being Poisson, rather than deleting them. Consider a single Bernoulli component $f^i_{t|t}(X)$ of the form (1), parameterised by $q^i_{t|t}$ and $f^i_{t|t}(x)$. We may choose to approximate this component as being Poisson:

$$\tilde{f}^i_{t|t}(X) = \exp(-\Lambda^i_{t|t}) \prod_{x \in X} \lambda^i_{t|t}(x)$$

where $\Lambda^i_{t|t} = \int \lambda^i_{t|t}(x) \mathrm{d}x$, and we define $\tilde{f}^i_{t|t}(x) \triangleq \lambda^i_{t|t}(x)/\Lambda^i_{t|t}$ for later use. The distortion caused by this approximation may be measured by the multi-target KL divergence: [1, p513]

$$D(f^i_{t|t}(X) || \tilde{f}^i_{t|t}(X)) = \int f^i_{t|t}(X) \log \frac{f^i_{t|t}(X)}{\tilde{f}^i_{t|t}(X)} \delta X$$

$$= (1 - q^i_{t|t}) \log \frac{1 - q^i_{t|t}}{\exp(-\Lambda^i_{t|t})} +$$

---
[5] The multi-target Bernoulli components are not naturally independent but they are *approximated* as such [4], as in the MeMBer [1,3].

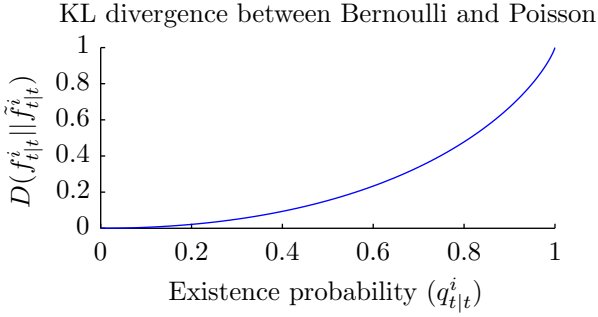

Fig. 4. Multi-target KL divergence between multi-target Bernoulli distribution and best-fit Poisson distribution as a function of target existence probability $q_{t|t}^i$.

$$+ \int q_{t|t}^i f_{t|t}^i(x) \log \frac{q_{t|t}^i f_{t|t}^i(x)}{\exp(-\Lambda_{t|t}^i)\Lambda_{t|t}^i \tilde{f}_{t|t}^i(x)} dx$$
$$= (1-q_{t|t}^i)\log(1-q_{t|t}^i) + (1-q_{t|t}^i)\Lambda_{t|t}^i +$$
$$+ q_{t|t}^i[\log q_{t|t}^i + \Lambda_{t|t}^i - \log \Lambda_{t|t}^i] + q_{t|t}^i D(f_{t|t}^i(x)\|\tilde{f}_{t|t}^i(x))$$

As shown in [1, p579], it is optimal to set

$$\tilde{f}_{t|t}^i(x) = f_{t|t}^i(x), \quad \Lambda_{t|t}^i = q_{t|t}^i, \quad \lambda_{t|t}^i(x) = q_{t|t}^i f_{t|t}^i(x) \quad (11)$$

The value of the KL divergence at this optimal choice is:

$$D(f_{t|t}^i(X)\|\tilde{f}_{t|t}^i(X)) = q_{t|t}^i + (1-q_{t|t}^i)\log(1-q_{t|t}^i)$$

The value of this KL divergence is shown in Fig. 4 as a function of the target existence probability $q_{t|t}^i$. The figure demonstrates that the distortion caused by the approximation (as measured by the KL divergence) is very small for existence probabilities less than 0.1 (when $q_{t|t}^i = 0.1$ we have $D(f_{t|t}^i(X)\|\tilde{f}_{t|t}^i(X)) \approx 0.005$, and when $q_{t|t}^i = 0.2$ we have $D(f_{t|t}^i(X)\|\tilde{f}_{t|t}^i(X)) \approx 0.02$).

Theorem 1 shows that the KL divergence between the overall multi-target distribution comprised of independent components (*e.g.*, (10)) and a modified multi-target distribution in which approximations have been made to a number of these components is bounded by the sum of the KL divergences between the components and their respective approximations.[6]

**Theorem 1.** *Let*

$$f(X) = \sum_{W \subseteq X} g(W)h(X-W)$$
$$\tilde{f}(X) = \sum_{W \subseteq X} \tilde{g}(W)\tilde{h}(X-W)$$

*Then* $D(f\|\tilde{f}) \leq D(g\|\tilde{g}) + D(h\|\tilde{h})$.

The proof of this theorem is in the appendix. Thus the overall distortion we cause to the complete multi-target distribution by approximating a number of components is bounded by the sum of the individual component distortions, which depend only on the component existence probabilities. Accordingly, we may approximate the components with lowest existence probabilities such that the sum of the distortions is less than an overall distortion budget.

[6]The theorem is stated and proven for the two component case; the general case follows simply by induction.

When any number of Bernoulli tracks are approximated as being Poisson, the resulting multi-target distribution is equivalent to one in which those tracks are dropped, and their intensity is added onto the undetected target intensity. To confirm this, denote the subset of tracks that we retain as $\tilde{\mathcal{T}}_t$; the approximated distribution is then

$$\tilde{G}_{t|t}[h] \propto \exp\{\lambda_{t|t}^u[h]\} \cdot \prod_{i \in \mathcal{T}_t \setminus \tilde{\mathcal{T}}_t} \tilde{G}_{t|t}^i[h] \cdot \prod_{i \in \tilde{\mathcal{T}}_t} G_{t|t}^i[h]$$
$$\propto \exp\{\lambda_{t|t}^u[h]\} \cdot \prod_{i \in \mathcal{T}_t \setminus \tilde{\mathcal{T}}_t} \exp\{\lambda_{t|t}^i[h]\} \cdot \prod_{i \in \tilde{\mathcal{T}}_t} G_{t|t}^i[h]$$
$$= \exp\left\{\lambda_{t|t}^u[h] + \sum_{i \in \mathcal{T}_t \setminus \tilde{\mathcal{T}}_t} \lambda_{t|t}^i[h]\right\} \cdot \prod_{i \in \tilde{\mathcal{T}}_t} G_{t|t}^i[h]$$
$$= \exp\{\tilde{\lambda}_{t|t}^u[h]\} \cdot \prod_{i \in \tilde{\mathcal{T}}_t} G_{t|t}^i[h]$$

where

$$\tilde{\lambda}_{t|t}^u[h] \triangleq \lambda_{t|t}^u[h] + \sum_{i \in \mathcal{T}_t \setminus \tilde{\mathcal{T}}_t} \lambda_{t|t}^i[h]$$

or, in intensity form,

$$\tilde{\lambda}_{t|t}^u(x) \triangleq \lambda_{t|t}^u(x) + \sum_{i \in \mathcal{T}_t \setminus \tilde{\mathcal{T}}_t} \lambda_{t|t}^i(x)$$

We refer to this concept as *recycling*, since the tracks that we delete are re-used in generation of new tracks in subsequent measurement scans. Not surprisingly, it can be shown that if the prior distribution is purely Poisson (*i.e.*, there are no pre-existing target tracks) and we choose to recycle all posterior tracks, then the posterior Poisson distribution is equivalent to that obtained using the PHD. This can be easily seen by observing that, if there are no prior tracks, $p^i(a) = 1$, hence by (9) and (11),

$$\lambda_{t|t}^i(x) = \frac{f_t(z|x)P_t^d(x)\lambda_{t|t-1}^u(x)}{\lambda_t^{fa}(z) + \lambda_{t|t-1}^u[f_t(z|\cdot)P_t^d]}$$

Combining this with (6) and setting $\tilde{\mathcal{T}}_t = \emptyset$ (*i.e.*, recycling all tracks), we obtain an update for a measurement set $Z_t = \{z_t^1, \ldots, z_t^{m_t}\}$ of

$$\tilde{\lambda}_{t|t}^u(x) = [1 - P_t^d(x)]\lambda_{t|t-1}^u(x) +$$
$$+ \sum_{j=1}^{m_t} \frac{f_t(z_t^j|x)P_t^d(x)\lambda_{t|t-1}^u(x)}{\lambda_t^{fa}(z_t^j) + \lambda_{t|t-1}^u[f_t(z_t^j|\cdot)P_t^d]}$$

which is the PHD update [2].

By recycling a subset of tracks, we permit the large mass of tracks with low probability of existence to be represented efficiently by the Poisson distribution, while maintaining explicit Bernoulli tracks on the subset with non-negligible probability of detection. Representing low probability of existence tracks via the Poisson distribution reduces the computational burden due to data association. Furthermore, if the Poisson distribution is represented as a discrete grid, then there is no computational cost associated with representing additional tracks.

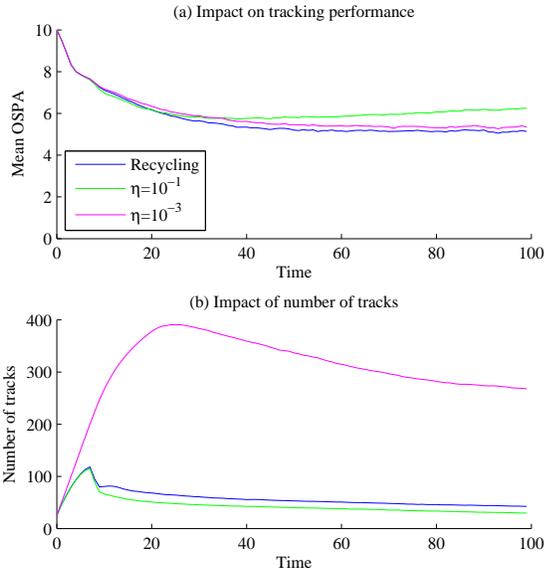

Fig. 5. Performance of recycling-based method, compared to alternatives. Top figure shows tracking performance measured by MOSPA, while bottom figure shows average number of tracks maintained.

In practice, this approximation allows the system to gradually accrue confidence in the presence of a target before choosing to maintain an explicit track. As shown in (7), the existence probability of a new, isolated track is the ratio between the undetected target intensity (in the vicinity of the measurement) and the undetected target intensity plus the false alarm intensity, so the intensity added by recycling will cause the existence probability of a new track due to a later measurement in the same vicinity to be increased, reducing the likelihood that the track will again be recycled.

### B. Example: Reduction of tentative tracks

We demonstrate the utility of recycling using the same scenario as Section III-A, and the same birth intensity and discretisation as Section III-B. Following the result in Fig. 4, we project tracks with a probability of existence $q_{t|t}^i < 10^{-1}$ onto the Poisson component (*i.e.*, we recycle them). We compare the performance of the method utilising recycling to alternatives which delete tracks with $q_{t|t}^i < \eta$, where $\eta = 10^{-1}$ or $\eta = 10^{-3}$ (the latter value was used in the simulations in Sections III-A and III-B).

Fig. 5 demonstrates the utility of the recycling. The top diagram shows the tracking performance, measured by the MOSPA metric. The method utilising recycling achieves similar (and even slightly improved) performance to the pure multi-Bernoulli method with $\eta = 10^{-3}$. The performance of the multi-Bernoulli method with $\eta = 10^{-1}$ reduces through the simulation as the undetected target intensity lowers, and the tracker becomes unable to initiate tracks on new targets (as they are deleted before being able to be confirmed).

The practical advantage of recycling is in the number of multi-Bernoulli components that need to be maintained. Fig. 5(b) shows the average number of components maintained by each method versus time. The method utilising recycling maintains slightly more tracks than the method with $\eta = 10^{-1}$, or around a quarter of the tracks of the method with $\eta = 10^{-3}$. Thus recycling permits similar performance to the alternative with a very low track deletion threshold, using only a fraction of the number of tracks (and, in turn, a reduction in computation for gating, data association, *etc*).

The result is somewhat surprising since the discretisation used for the Poisson component was quite coarse, *e.g.*, the measurement variance is 1, yet the discretisation resolution is 4. Thus the method appears to be quite robust to the discretisation used in a grid representation.

Finally, we note that the grid representation is not necessary for the algorithm; indeed, a sample-based representation is also possible (as in [7]), as is a hybrid approximation (*e.g.*, retaining Gaussian mixture components until their variance grows sufficiently to be adequately represented by a grid).

## V. RELATED WORK

The landmark paper [12] incorporated a Poisson distribution of undetected targets. Its primary role was to provide a distribution of newly detected targets for multiple sensors with overlapping fields of view, and as such, it was approximated as being uniform at the beginning of each scan.[7] The RFS framework [1] has produced an improved set of analytical tools for Bayesian modelling in tracking problems, and the birth, transition and death models we employ are standard in the PHD. Despite this resurgence, the concept of maintaining a distribution of undetected targets appears not to have reentered the mainstream. The Gaussian mixture and sample-based representations commonly used to represent the PHD generally struggle to adequately model the diffuse distribution of undetected targets.

To the author's knowledge, the concept of mixing multi-Bernoulli and PHD representations was first proposed simultaneously in the pre-print [6] and in [7]. The latter work was somewhat heuristic, in that it was not supported by the RFS-based derivation in [4,6]. This, for example, resulted in retaining the standard hypothesis weight for a target not being associated to any other measurement, *e.g.*, $w_{t|t}^{i,a} = \lambda_t^{\text{fa}}(z)$, as opposed to the value in the previous section, which results from the RFS derivation. The analytical tools provided by the RFS framework also permit an analysis of the error cause by a Poisson approximation to a multi-Bernoulli component (*i.e.*, Fig. 4), which permits a system designer to set the threshold for when to use the approximation.

A significant contribution in [7] was the development of a sophisticated sample-based method for representing the PHD intensity. A surprising outcome of the experiment in

---

[7]Presumably, this is due to the computational limitations of the day.

Section IV-B is that a grid-based approximation of the intensity with a resolution several times lower than the sensor accuracy still appears to be adequate to achieve equivalent performance to a system using a very low track deletion threshold.

## VI. CONCLUSION

This paper has shown the practical advantage of hybrid tracking systems that utilise both a Poisson component and a multi-Bernoulli component. Our sister paper [4] showed that this is a natural result of the derivation of the Bayes RFS filter, and that the Poisson component represents the intensity of targets that remain undetected. Here we have demonstrated the practical advantage of maintaining such a quantity, improving track initiation performance in dynamic environments. Subsequently, we derived the method of recycling, which additionally utilises the Poisson component to represent Bernoulli components with low probability of existence, and demonstrated its practical utility in tracking problems.

## APPENDIX
## SUB-ADDITIVITY OF KL DIVERGENCE

This section presents the proof of Theorem 1, which in turn requires Lemma 1.

*Proof of Theorem 1:*

$$D(f||\tilde{f}) = \int \left[ \sum_{W \subseteq X} g(W)h(X - W) \right] \cdot$$
$$\cdot \log \frac{\left[ \sum_{W \subseteq X} g(W)h(X - W) \right]}{\left[ \sum_{W \subseteq X} \tilde{g}(W)\tilde{h}(X - W) \right]} \delta X$$
$$\overset{(a)}{\leq} \int \sum_{W \subseteq X} g(W)h(X - W) \cdot$$
$$\cdot \log \frac{g(W)h(X - W)}{\tilde{g}(W)\tilde{h}(X - W)} \delta X$$
$$= \int \sum_{W \subseteq X} g(W)h(X - W) \log \frac{g(W)}{\tilde{g}(W)} \delta X$$
$$+ \int \sum_{W \subseteq X} g(W)h(X - W) \log \frac{h(X - W)}{\tilde{h}(X - W)} \delta X$$
$$\overset{(b)}{=} D(g||\tilde{g}) + D(h||\tilde{h})$$

where $(a)$ is a consequence of the log-sum inequality, [13, p29], and $(b)$ is the result of Lemma 1. ∎

**Lemma 1.**
$$\int \sum_{W \subseteq X} a(W)b(X - W) \delta X = \int a(W) \delta W \cdot \int b(Y) \delta Y$$

*Proof:* For simplicity, when $n = 0$ let $\int \cdots \int a(\{x_1, \ldots, x_n\}) \mathrm{d}x_1 \cdots \mathrm{d}x_n \triangleq a(\emptyset)$. Then:

$$\int \sum_{W \subseteq X} a(W)b(X - W) \delta X$$
$$\triangleq \sum_{n=0}^{\infty} \frac{1}{n!} \int \cdots \int \sum_{W \subseteq \{x_1, \ldots, x_n\}} a(W) \cdot$$
$$\cdot b(\{x_1, \ldots, x_n\} - W) \mathrm{d}x_1 \cdots \mathrm{d}x_n$$
$$= \sum_{n=0}^{\infty} \sum_{I \subseteq \{1, \ldots, n\}} \frac{1}{n!} \int \cdots \int a(\{x_i | i \in I\}) \cdot$$
$$\cdot b(\{x_i | i \in \{1, \ldots, n\} - I\}) \mathrm{d}x_1 \cdots \mathrm{d}x_n$$
$$\overset{(a)}{=} \sum_{n=0}^{\infty} \sum_{m=0}^{n} \frac{1}{n!} \cdot \frac{n!}{m!(n-m)!} \cdot$$
$$\cdot \int \cdots \int a(\{x_1, \ldots, x_m\}) \mathrm{d}x_1 \cdots \mathrm{d}x_m \cdot$$
$$\cdot \int \cdots \int b(\{x_{m+1}, \ldots, x_n\}) \mathrm{d}x_{m+1} \cdots \mathrm{d}x_n$$
$$\overset{(b)}{=} \sum_{m=0}^{\infty} \sum_{l=0}^{\infty} \frac{1}{m!} \int \cdots \int a(\{x_1, \ldots, x_m\}) \mathrm{d}x_1 \cdots \mathrm{d}x_m \cdot$$
$$\cdot \frac{1}{l!} \int \cdots \int b(\{x_1, \ldots, x_l\}) \mathrm{d}x_1 \cdots \mathrm{d}x_l$$
$$= \int a(W) \delta W \cdot \int b(Y) \delta Y$$

where $(a)$ results from the observation that terms in the sum over $I$ with the same number of elements in $I$ will have the same value, and $(b)$ results from letting $l = (n - m)$ and reordering terms in the two-dimensional summation. ∎

## REFERENCES

[1] R. P. S. Mahler, *Statistical Multisource-Multitarget Information Fusion*. Norwood, MA: Artech House, 2007.
[2] ——, "Multitarget Bayes filtering via first-order multitarget moments," *IEEE Trans. Aerosp. Electron. Syst.*, vol. 39, no. 4, pp. 1152–1178, Oct. 2003.
[3] B.-T. Vo, B.-N. Vo, and A. Cantoni, "The cardinality balanced multi-target multi-Bernoulli filter and its implementations," *IEEE Trans. Signal Process.*, vol. 57, no. 2, pp. 409–423, Feb. 2009.
[4] J. L. Williams, "Alternative multi-Bernoulli filters," in *Submitted to Proc. 15th International Conference on Information Fusion*, July 2012. [Online]. Available: http://arxiv.org/
[5] ——, "Graphical model approximations of random finite set filters," arXiv, e-print arXiv:1105.3298v2, August 2011. [Online]. Available: http://arxiv.org/abs/1105.3298
[6] ——, "Search theory approaches to radar resource allocation," in *Proc. 7th U.S./Australia Joint Workshop on Defense Applications of Signal Processing (DASP)*, Coolum, Australia, July 2011.
[7] P. Horridge and S. Maskell, "Using a probabilistic hypothesis density filter to confirm tracks in a multi-target environment," in *Proc. 6th Workshop on Sensor Data Fusion*, Berlin, Germany, October 2011.
[8] J. L. Williams and R. A. Lau, "Convergence of loopy belief propagation for data association," in *Sixth International Conference on Intelligent Sensors, Sensor Networks and Information Processing*, Brisbane, Australia, December 2010.
[9] P. O. Vontobel, "The Bethe permanent of a non-negative matrix," arXiv, e-print arXiv:1107.4196v1, July 2011. [Online]. Available: http://www.arxiv.com/abs/1107.4196
[10] D. Schuhmacher, B.-T. Vo, and B.-N. Vo, "A consistent metric for performance evaluation of multi-object filters," *IEEE Trans. Signal Process.*, vol. 56, no. 8, pp. 3447–3457, 2008.
[11] L. D. Stone, C. A. Barlow, and T. L. Corwin, *Bayesian multiple target tracking*. Boston: Artech House, 1999.
[12] S. Mori, C.-Y. Chong, E. Tse, and R. Wishner, "Tracking and classifying multiple targets without a priori identification," *IEEE Trans. Autom. Control*, vol. 31, no. 5, pp. 401–409, May 1986.
[13] T. M. Cover and J. A. Thomas, *Elements of Information Theory*. New York, NY: John Wiley and Sons, 1991.